\documentclass{article}[11pt]
\usepackage{a4wide}
\usepackage{stmaryrd,latexsym,bm,bbm}
\usepackage{epsfig,graphics}
\usepackage{amsmath,amssymb,amsfonts}
\usepackage{color}
\usepackage{epstopdf}
\usepackage{slashed}

\usepackage[font=footnotesize, labelfont=bf]{caption,subcaption}
\usepackage[square,comma,sort&compress,numbers]{natbib}
\usepackage[
colorlinks=true,
linkcolor=black,
breaklinks=true,
urlcolor=blue,
citecolor=green]{hyperref}
\usepackage[utf8]{inputenc}
\usepackage{booktabs}
\usepackage{dcolumn}
\usepackage{makecell}
\usepackage{multirow}

\definecolor{darkblue}{rgb}{0.,0.,0.7}
\definecolor{light-blue}{rgb}{0.8,0.85,1}
\definecolor{green}{rgb}{0,0.6,0}
\definecolor{blueviolet}{rgb}{0.541, 0.169, 0.886}
\definecolor{fuchsia}{rgb}{1.0, 0, 1.0}

\newcommand{\Lag}{\mathcal{L}}
\newcommand{\eps}{\epsilon}

\newcommand{\mev}{\mathrm{MeV}}
\newcommand{\gev}{\mathrm{GeV}}

\newcommand{\beq}{\begin{equation}}
\newcommand{\eeq}{\end{equation}}
\newcommand{\beqa}{\begin{eqnarray}}
\newcommand{\eeqa}{\end{eqnarray}}

\begin{document}

\title{Decay behavior of the strange and beauty partners of $P_c$ hadronic molecules}
\date{\today}
\author{Yong-Hui Lin$^{1,2,}$\footnote{Email address:
      \texttt{linyonghui@itp.ac.cn} }~ ,
      Chao-Wei Shen$^{1,2,}$\footnote{Email address:
      \texttt{shencw@itp.ac.cn} }~  and
      Bing-Song Zou$^{1,2,3}$\footnote{Email address:
      \texttt{zoubs@itp.ac.cn} }
        \\[2mm]
      {\it\small$^1$CAS Key Laboratory of Theoretical Physics, Institute
      of Theoretical Physics,}\\
      {\it\small  Chinese Academy of Sciences, Beijing 100190,China}\\
      {\it\small$^2$University of Chinese Academy of Sciences (UCAS), Beijing 100049, China} \\
      {\it\small$^3$Synergetic Innovation Center for Quantum Effects and Applications (SICQEA)},\\
      {\it\small Hunan Normal University, Changsha 410081, China} \\
}

\maketitle

\begin{abstract}
We extend our previous study on the decay behavior of $P_c$ hadronic molecules to their strange and beauty partners.
While $P_c(4380)$ and $P_c(4450)$ locate just below the $\bar D\Sigma_c^*$ and $\bar D^*\Sigma_c$ thresholds, respectively,
their proposed strange partners $N(1875)$ and $N(2080)$ sit just below the $K\Sigma^*$ and $K^*\Sigma$ thresholds, respectively.  Using the effective Lagrangian approach as the same as for the study of $P_c$ hadronic molecular states and with the couplings determined by the $\mathrm{SU}(3)$ flavor symmetry, the decay patterns of $N(1875)$ and $N(2080)$ are obtained assuming them to be the $S$-wave $K\Sigma^*$ and $K^*\Sigma$ molecules, respectively. It is found that the measured decay properties of $N(1875)$ and $N(2080)$ are reproduced well. Our results suggest that both $N(1875)$ and $N(2080)$ can be ascribed as the hadronic molecular pentaquark states in the hidden strangeness sector with quantum number $I(J^P)=1/2({3/2}^-)$. Further checks on their hadronic molecule nature by various experiments are proposed. With the same approach, we also give our predictions for the decay behavior of the possible beauty partners of the $P_c$ hadronic molecular states. The $B^*\Lambda_b$ is found to be the largest decay mode for both $B\Sigma_b^*$ and $B^*\Sigma_b$ hadronic molecules, hence can be used to look for the hidden beauty pentaquark states in forthcoming experiments.
\end{abstract}

\medskip
\newpage

\section{Introduction} \label{sec:1}

There is a long history hunting for various pentaquark states~\cite{Guo:2017jvc,Chen:2016qju}. Some previous claimed ones faded away with time, while others got controversial interpretations.   Up to now, the most convincing evidence came from the observation of two hidden-charm pentaquark-like structures, $P_c^+(4380)$ and $P_c^+(4450)$ decaying to p-$J/\psi$, by LHCb collaboration~\cite{Aaij:2015tga} in 2015. The existence of such pentaquark states around this energy range has been predicted in
Refs.~\cite{Wu:2010jy,Wu:2010vk,Yang:2011wz,Yuan:2012wz,Xiao:2013yca}. The LHCb observation triggered
widespread further theoretical investigations on their nature~\cite{Guo:2017jvc,Chen:2016qju}. Especially, there is a very interesting fact about these two
states that the reported masses of $P_c^+(4380)$ and $P_c^+(4450)$ locate just below the thresholds of
$\bar{D}\Sigma_c^*$ and $\bar{D^*}\Sigma_c$ at 4382 MeV and 4459 MeV, respectively. This property seems strongly supporting the interpretation of $P_c^+(4380)$ and $P_c^+(4450)$ as the hadronic molecules composed of either $\bar{D}\Sigma_c^*$ or $\bar{D^*}\Sigma_c$ and inspired us to study their decay behavior under this molecular picture in our previous work~\cite{Lin:2017mtz}. As we claimed in that work, $P_c^+(4380)$ is more likely to be a spin-parity-${3/2}^-$
$\bar{D}\Sigma_c^*$ molecular state, while $P_c^+(4450)$ is a $\bar{D^*}\Sigma_c$ molecule with $J^P={5/2}^+$.
In this scenario, there may exist similar structures in strange and beauty sectors from quark flavor symmetry.
Thus, we extend our previous study of the $P_c$ states to their possible strange and beauty partners in the present work.

In the strange sector, there are indeed two $N^*$ states sit just below the corresponding $K\Sigma^*$ and $K^*\Sigma$ thresholds, {\sl i.e.},  $N(1875)$ and $N(2080)$, which were suggested to be the strange partners of the $P_c$ states~\cite{He:2017aps}. Possible existence of such hadronic molecules was also implicated in a study with the quark delocalization color screening model~\cite{Gao:2017hya}. We shall work out their decay patterns within the hadronic molecule picture and compare them with the experimental data in $\mathrm{PDG}$~\cite{Patrignani:2016xqp} to see whether they fit to be the strange partners of $P_c^+(4380)$ and $P_c^+(4450)$ or not. For the beauty sector, similar hadronic molecular pentaquark states were predicted to exist~\cite{Wu:2010rv,Xiao:2013jla} and a recent study with a unitary coupled-channel model~\cite{Shen:2017ayv} suggests there are more bound states with higher partial waves than in the case of hidden charm sector. We shall give predictions for their major decay modes to guide the future experimental search.

This paper is organized as follows: In Sec.~\ref{sec:2}, we introduce formalism and some details about the theoretical tools used to calculate the decay modes of exotic hadronic molecular states. In Sec.~\ref{sec:3}, the numerical results and discussion are presented.

\section{Formalism} \label{sec:2}
As listed in the latest Particle Date Group ($\mathrm{PDG}$) review~\cite{Patrignani:2016xqp}, the two-star nucleon
resonance $N(2080)$ in previous version has been split into a three-star $N(1875)$ and a two-star $N(2120)$ both with
spin-parity ${3/2}^-$. The mass and total decay width are claimed to be $1875\pm 20\ \mev$ and $250\pm 70\ \mev$, respectively, for
the $N(1875)$, while the values are $2120\pm 45\ \mev$ and $250\pm 130\ \mev$, respectively, for the $N(2120)$. A marked
feature of the hadronic molecules is that those molecular states should be very shallowly bounded~\cite{Guo:2017jvc}.
It means that in general the mass is slightly below the corresponding threshold in the hadronic molecular picture.
Since we will work in the $K\Sigma^*$ and $K^*\Sigma$ molecular pictures with their corresponding thresholds of 1880 MeV and 2086 MeV, respectively, we take the mass of $N(1875)$ as $1875\ \mev$
and the mass of $N(2120)$ as $2080\ \mev$ accordingly. For the latter we use its old name $N(2080)$ in the remaining part of this paper. Then these two exotic $N^*$ resonances are treated as an $S$-wave
$K\Sigma^*$ and $K^*\Sigma$ molecule, respectively. Based on this assumption, the decay modes of $N(1875)$ and $N(2080)$
can be obtained directly by means of the effective Lagrangian approach. Before that it should be mentioned that the
$S$-wave couplings of the exotic $N^*$s to the $K\Sigma^*$ or $K^*\Sigma$ channels can be estimated model-independently
with the Weinberg compositeness criterion. For the pure hadronic molecular case, it gets
that~\cite{Lin:2017mtz,Baru:2003qq,Weinberg:1965zz}
\beq
{g^2} = \frac{4\pi}{4 M m_2}  \frac{(m_1+m_2)^{5/2}} {(m_1 m_2)^{1/2}}
\sqrt{32\epsilon} , \label{eq:coupling}
\eeq
where $M$, $m_1$ and $m_2$ denote the masses of the exotic $N^*$s, $K(K^*)$ and $\Sigma({\Sigma}^*)$,
respectively, and $\eps$ is the binding energy. Assuming the physical state in question to be a pure $S$-wave hadronic
molecule, the relative uncertainty of the above approximation for the coupling constant is $\sqrt{2\mu\eps}\, r$ where
$\mu=m_1 m_2/(m_1+m_2)$ is the reduced mass of the bound particles, and $r$ is the range of forces which may be estimated
by the inverse of the mass of the particle that can be exchanged. Thus, for the $K\Sigma^*$ and $K^*\Sigma$ systems,
$r$ may be estimated as $1/m_{K^*}$ and $1/m_{K}$, respectively.

\begin{figure}[htbp]
\begin{center}
\includegraphics[width=0.5\textwidth]{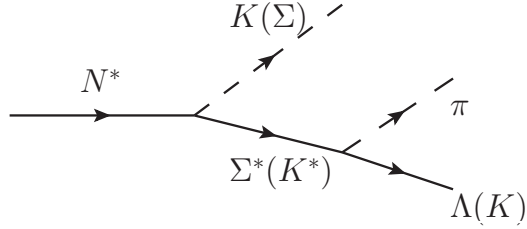}
\caption{The three-body decays of the exotic $N^*$s in the $K \Sigma^*$ and $K^*\Sigma$ molecular pictures.
\label{Fig:threebody}}
\end{center}
\end{figure}

\begin{figure}[htbp]
\begin{center}
\includegraphics[width=0.5\textwidth]{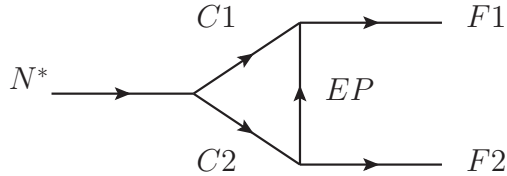}
\caption{The triangle diagram for the two-body decays of the exotic $N^*$s in the $K \Sigma^*$ and $K^*\Sigma$
molecular pictures, where $C1$, $C2$ denote the constituent particles of the composite system $K \Sigma^*$ and
$K^*\Sigma$, $F1$, $F2$ denote the final states, $EP$ denotes the exchanged particles.
\label{Fig:triangle}}
\end{center}
\end{figure}

Note that both $K^*$ and $\Sigma^*$ (which denotes the $\Sigma(1385)$ state in this work) are unstable with a sizable
decay width, while $K$ and $\Sigma$ are very stable. This leads that the significant three-body decays through the decay of
$K^*$ or $\Sigma^*$ must be considered in our case and the four-body decays through the decays of both two constituents
are strongly suppressed by the small widths of $K$ and $\Sigma$. The dominant three-body decays are given in
Fig.~\ref{Fig:threebody}, where the interactions between the final states have been neglected.
For the two-body decays, we take the same convention that used in our previous work~\cite{Lin:2017mtz}. To be concrete, we choose a
perturbative formula to provide a rough estimation for the total widths of $N(1875)$ and $N(2080)$ by calculating
the contributions from the relevant two-body decays, not the nonperturbative approach which may be more precise
to give the total widths for such a broad resonance. It is
\beq
{\rm d}\Gamma = \frac{F_I}{32 \pi^2} \overline{|{\cal M}|^2}
\frac{|\mathbf{p_1}|}{M^2} {\rm d}\Omega,
\label{eq:widths}
\eeq
where ${\rm d}\Omega = {\rm d}\phi_1 {\rm d}(\cos{\theta_1})$ is the solid angle of particle 1, $M$ is the mass
of the initial $N^*$, the factor $F_I$ is from the isospin symmetry, and the
polarization-averaged squared amplitude $\overline{|{\cal M}|^2}$ means $\frac14 \sum_\text{spin} |{\cal M}|^2$.
The two-body decays can be described as the typical triangle diagram shown in Fig.~\ref{Fig:triangle}.
And all the two-body decay modes included in our calculation are listed in Table~\ref{Tab:modes}.
\begin{table}[htpb]
\centering
\caption{\label{Tab:modes}All possible decay channels for the $N(1875)$ and $N(2080)$.}
\begin{tabular}{c|*{2}{c}}
\toprule\morecmidrules\toprule
\thead{Initial state} & \thead{Final states} & \thead{Exchanged particles} \\
\Xhline{0.8pt}
\multirow{5}*{$N(1875)(K \Sigma^*)$} & $\omega p$, $\rho N$ & $K$, $K^*$ \\
\Xcline{2-3}{0.4pt}
& $K\Sigma$, $K\Lambda$ & $\rho$ \\
\Xcline{2-3}{0.4pt}
& $\sigma p$ & $K$ \\
\Xcline{2-3}{0.4pt}
& $\eta p$, $\pi\Delta$ & $K^*$ \\
\Xcline{2-3}{0.4pt}
& $\pi N$ & $K^*$, $\Lambda$, $\Sigma$ \\
\Xhline{0.8pt}
\multirow{3}*{$N(2080)(K^* \Sigma)$} & $K^*\Lambda$, $K\Sigma$, $K\Lambda$, $K\Sigma^*$ & $\rho$, $\pi$ \\
\Xcline{2-3}{0.4pt}
& $\omega p$, $\rho p$, $\phi p$, $\eta p$, $\pi\Delta$ & $K^*$, $K$ \\
\Xcline{2-3}{0.4pt}
& $\pi N$ & $K^*$, $K$, $\Lambda$, $\Sigma$ \\
\Xcline{2-3}{0.4pt}
& $K\Lambda(1520)$, $K\Lambda(1405)$ & $\pi$ \\
\bottomrule\morecmidrules\bottomrule
\end{tabular}
\end{table}
The effective Lagrangians which describe the vertices that appear in the diagrams above and the determination
of the needed coupling constants are discussed below. First of all, we adopt
the Lorentz covariant orbital-spin scheme proposed in Ref.~\cite{Zou:2002yy} for the $S$-wave interactions between
the initial $N^*$ states and $K \Sigma^*$ or $K^* \Sigma$ channels. The Lagrangians are given as
\beqa
\Lag_{K \Sigma^* N(1875)} &=& g_{N^* K \Sigma^*}^{} \bar N^*_{\mu} K \Sigma^{* \mu} + H.c. , \notag \\
\Lag_{K^* \Sigma N(2080)} &=& g_{N^* K^* \Sigma}^{} \bar N^*_{\mu} K^{* \mu} \Sigma + H.c. .
\label{eq:vertex0}
\eeqa
As mentioned above, these two $S$-wave couplings, $g_{N^* K \Sigma^*}^{}$ and $g_{N^* K^* \Sigma}^{}$, can be
determined directly by the Eq.~\eqref{eq:coupling}.
Note that for the $S$-wave $K^* \Sigma$ two-body system, another possible quantum number assignment is ${1/2}^-$.
Then we also analyze the properties of spin-parity-${1/2}^-$ $K^* \Sigma$
molecular state in this work. The mass of this state was chosen as the same as the spin-parity-${3/2}^-$ one. This
$S$-wave interaction is described as
\beq
\Lag_{K^* \Sigma N(2080)}^\prime = g_{N^* K^* \Sigma}^{\prime} \bar N^* \gamma_5 \tilde{\gamma}_{\mu}K^{* \mu} \Sigma
 + H.c. ,
\eeq
where $\tilde{\gamma}_{\mu}=-\left(-g_{\mu\nu}+\frac{p_\mu p_\nu}{p^2}\right) \gamma^\nu$ with $p_\mu$ the
momentum of initial $N^*$ state. In the same way, the coupling constant $g_{N^* K^* \Sigma}^{\prime}$ can
also be inferred from the Weinberg compositeness criterion. For the other vertices in the above two-body
and three-body decays, most of them are exactly the same with those in our previous work. Here for simplicity we
just list some effective Lagrangians for those vertices that have not appeared before. This includes the
$KK\sigma$, $K^*\Sigma^*\Delta$, $\Sigma\pi\Lambda(1520)$ and $\Sigma\pi\Lambda(1405)$ vertices that are described
by the following Lagrangians respectively~\cite{Schutz:1994ue,Ronchen:2012eg,Zou:2002yy}.
\beqa
\Lag_{KK\sigma} &=& g_{KK\sigma}^{} \partial_\mu \bar K^\dag \partial^\mu K \sigma , \notag \\
\Lag_{K^*\Sigma^*\Delta} &=& g_{K^*\Sigma^*\Delta}^{} \bar \Sigma^*_{\tau}(\gamma^\mu-i \frac{\kappa_{K^*\Sigma^*\Delta}}
{m_{\Sigma^*}+m_\Delta} \sigma^{\mu\nu}\partial_\nu) K^*_\mu \Delta^{\tau} + H.c. , \notag \\
\Lag_{\Sigma\pi\Lambda(1520)} &=& g_{\Lambda(1520)\pi\Sigma}^{} \bar \Sigma \gamma^5 \gamma^\mu \partial_\mu \partial^\nu
\pi {\Lambda(1520)}_{\nu} + H.c. , \notag \\
\Lag_{\Sigma\pi\Lambda(1405)} &=& g_{\Lambda(1405)\pi\Sigma}^{} \bar \Sigma \pi \Lambda(1405) + H.c. .
\label{eq:vertex-else}
\eeqa
Besides, it should be mentioned that a momentum-carried pseudoscalar-baryon-baryon interaction is adopted for the
vertices $KN\Sigma$ and $KN\Lambda$ in the $\pi N$ channel of $N(1875)$. The corresponding Lagrangian is
\beq
\Lag_{P B_1 B_2} = g_{P B_1 B_2}^{} \bar B_1 \gamma^5 \gamma^\mu \partial_\mu P B_2 ,
\label{eq:vertex-PBB}
\eeq
where $g_{P B_1 B_2}^{}$ is usually substituted by $f_{P B_1 B_2}^{}/m_P$. It is a little different from what happened
in Ref.~\cite{Lin:2017mtz}. In that work, we take the non-momentum formula for this kind vertex since the involved pseudoscalars
are almost on their mass shell. While for the $K\Sigma^*$ molecule,
the large mass difference between constituent $K$ and $\Sigma^*$ leads the $\Sigma^*$ being almost on its mass shell with $K$
must be off shell to some extent. Then the momentum in the interaction needs to be kept to include this off-shell
contribution. Now let us focus on the couplings needed for the estimation of the partial decay widths. The coupling
constants $g_{\Lambda(1520)\pi\Sigma}^{}$ and $g_{\Lambda(1405)\pi\Sigma}^{}$ are deduced from the experimental data
of the partial decay widths of $\Lambda(1520)$ and $\Lambda(1405)$ into $\pi\Sigma$ channel respectively. And the
coupling $g_{KK\sigma}$ is related to the $g_{\pi\pi\sigma}$ with the $\mathrm{SU(3)}$ flavor
symmetry~\cite{deSwart:1963pdg}. It happens to be $g_{KK\sigma}=\sqrt{2} g_{\pi\pi\sigma}$ and $g_{\pi\pi\sigma}=
4.59\ \mathrm{GeV}^{-1}$ taken from Ref.~\cite{He:2015yva,Krehl:1997kg,Schutz:1994ue}. All other effective
coupling constants are also determined under the $\mathrm{SU(3)}$ flavor symmetry scheme. For the detailed description
of this scheme, we refer the interesting readers to the related papers~\cite{deSwart:1963pdg,Ronchen:2012eg,Doring:2010ap}.
Here we summarize the exact values of the involved parameters in the $\mathrm{SU(3)}$ relations and other coupling
constants into the Table~\ref{table:constants}.
\begin{table}[htpb]
\centering
\caption{\label{table:constants}The parameters from the $\mathrm{SU(3)}$ flavor symmetry and other coupling
constants used in the present work. These parameters are taken from
Refs.~\cite{Wu:2010vk,Doring:2010ap,Janssen:1996kx}. The $P$, $V$, $B$ and $D$ denote the pseudoscalar,
vector mesons, octet and decuplet baryons respectively. Only absolute values of the couplings are listed with
their signs ignored.}
\scalebox{0.85}{
\begin{tabular}{*{11}{c}}
			\toprule\morecmidrules\toprule
			$\alpha_{BBP}$ & $\alpha_{BBV}$ & \thead{$g_{BBP}$ \\ $(\mathrm{GeV}^{-1})$} & $g_{BBV}$ & $g_{VPP}$
			& \thead{$g_{VVP}$ \\ $(\mathrm{GeV}^{-1})$} & \thead{$g_{PBD}$ \\ $(\mathrm{GeV}^{-1})$} & \thead{$g_{VBD}$ \\
				$(\mathrm{GeV}^{-1})$}  & \thead{$g_{PDD}$ \\ $(\mathrm{GeV}^{-1})$} & $g_{VDD}$ & $\kappa_{VDD}$ 	\\
			\Xhline{0.4pt}
			0.4 & 1.0 & 7.06 & 3.25 & 3.02 & 12.84 & 15.19 & 20.68 & 12.71 & 7.67 & 6.1 	\\
			\Xhline{0.8pt}
			$g_{KK\omega}$ & \thead{$g_{K^*K\omega}$ \\ $(\mathrm{GeV}^{-1})$} & $g_{KK\rho}$ & \thead{$g_{K^*K\rho}$ \\
				$(\mathrm{GeV}^{-1})$} & $g_{K^*K\eta}$ & \thead{$g_{K^*K^*\eta}$ \\ $(\mathrm{GeV}^{-1})$} & $g_{K^*K\pi}$
			& \thead{$g_{K^*K^*\pi}$ \\ $(\mathrm{GeV}^{-1})$} & $g_{K^*K^*\rho}$ & $g_{K^*K^*\omega}$ & $g_{K^*K^*\phi}$	\\
			\Xhline{0.4pt}
			3.02 & 6.42 & 3.02 & 6.42 & 5.23 & 11.12 & 3.02 & 6.42 & 3.02 & 3.02 & 4.27	\\
			\Xhline{0.8pt}
			\thead{$g_{KN \Sigma^*}$ \\ $(\mathrm{GeV}^{-1})$} & \thead{$g_{K^*N \Sigma^*}$ \\ $(\mathrm{GeV}^{-1})$}
			& \thead{$g_{\rho \Sigma \Sigma^*}$ \\ $(\mathrm{GeV}^{-1})$} & \thead{$g_{\rho \Lambda \Sigma^*}$ \\
				$(\mathrm{GeV}^{-1})$} & $g_{K^* \Sigma^* \Delta}$ & \thead{$g_{KN \Sigma}$ \\ $(\mathrm{GeV}^{-1})$}
			& \thead{$g_{KN \Lambda}$ \\ $(\mathrm{GeV}^{-1})$} & \thead{$g_{\pi \Sigma \Sigma^*}$ \\ $(\mathrm{GeV}^{-1})$}
			& \thead{$g_{\pi \Lambda \Sigma^*}$ \\ $(\mathrm{GeV}^{-1})$} & $g_{\rho \Sigma \Sigma}$	
			& $g_{\rho \Lambda \Sigma}$ \\
			\Xhline{0.4pt}
			4.38 & 5.97 & 5.97 & 14.62 & 6.26 & 1.91 & 9.92 & 4.38 & 14.63 & 4.60 & 0.0 	\\
			\Xhline{0.8pt}
			\thead{$g_{\pi \Sigma \Sigma}$ \\ $(\mathrm{GeV}^{-1})$} & \thead{$g_{\pi \Lambda \Sigma}$ \\
				$(\mathrm{GeV}^{-1})$} & $g_{K^*N \Sigma}$ & $g_{K^*N \Lambda}$ & \thead{$g_{K^*\Sigma\Delta}$ \\
				$(\mathrm{GeV}^{-1})$} & \thead{$g_{K\Sigma\Delta}$ \\ $(\mathrm{GeV}^{-1})$} \\
			\Xhline{0.4pt}
			7.64 & 9.35 & 2.30 & 3.98 & 16.89 & 12.40	\\
\bottomrule\morecmidrules\bottomrule 	
\end{tabular}
}
\end{table}

Finally, there is a technical issue appearing when we get to calculate the triangle diagrams with the
effective Lagrangians above. Some of the amplitudes, corresponding to the exchange of a pseudoscalar meson
for the $D$-wave decay modes~\cite{Albaladejo:2015dsa, Shen:2016tzq}, are ultraviolet (UV) finite while the
others diverge. Nevertheless, even the UV finite loops receive short-distance contributions when we integrate
over the whole momentum space. As well known, there are several kinds of approaches to deal with the UV
divergence problems, such as the dimensional regularization, the momentum cut-off regularization and the form
factor method. Here we adopt the last strategy and employ the following Gaussian regulator to suppress
short-distance contributions and thus can render all the amplitudes UV
finite~\cite{Guo:2017jvc,Epelbaum:2008ga,Guo:2013sya,Guo:2013xga,HidalgoDuque:2012pq,Nieves:2011vw,
Nieves:2012tt,Valderrama:2012jv},
\beq
f(\bm{p}^2 /\Lambda_0^2) = {\rm{exp}}(-\bm{p}^2 /\Lambda_0^2),
\label{eq:regualtor}
\eeq
where $\bm p$ is the spatial part of the loop momentum and $\Lambda_0$ is an ultraviolet cut-off.
The cutoff $\Lambda_0$ denotes a hard momentum scale which suppresses the contribution of the two
constituents at short distances $\sim 1/\Lambda_0$. There is no universal criterion for choosing
the cut-off, but as a general rule the value of $\Lambda_0$ should be much larger than the typical
momentum in the bound state, given by $\sqrt{2\mu\epsilon}$. And it should also not be too large
since we have neglected all other degrees of freedom, except for the two constituents, which would
play a role at short distances. Here we range $\Lambda_0$ from $0.6 \ \mathrm{GeV}$ to $1.4\ \mathrm{GeV}$. In addition,
as described in our previous work a usual form factor chosen as Eq.~\eqref{eq:ff} is also introduced
to suppress the off-shell contributions for the exchanged particles.
\beq
f(q^2) = \frac{\Lambda_1^4}{(m^2 - q^2)^2 + \Lambda_1^4},
\label{eq:ff}
\eeq
where $m$ is the mass of the exchanged particle and $q$ is the corresponding momentum. The cut-off $\Lambda_1$ varies
from $0.8\ \mathrm{GeV}$ to $2.0\ \mathrm{GeV}$.

For the hidden beauty case, the same approach is used with all the parameters obtained through heavy quark symmetry from our study of $P_c$ states~\cite{Lin:2017mtz} except for differences between the masses of charm hadrons and beauty hadrons.

\section{Results and Discussions} \label{sec:3}
With the aforementioned ingredients, the partial decay widths of $N(1875)$ and $N(2080)$ to
the channels listed in Table~\ref{Tab:modes} are calculated. It should be mentioned that there are two major
sources of uncertainty in our model, {\sl i.e.}, the determination of the coupling constants and the choice of cutoffs
$\Lambda_0$ and $\Lambda_1$. These are also the general flaws of the effective field theory.
In our case, the $\mathrm{SU(3)}$ flavor symmetry should definitely be broken more or less due to the mass
differences among $u$, $d$ and $s$ quarks. It may lead to some deviations from the coupling constants
we obtain by assuming SU(3) flavor symmetry. Nevertheless, the $\mathrm{SU(3)}$ flavor symmetry still
can give a basic estimation for the couplings appearing in the effective  Lagrangians.
Based on this approximation, a rough estimate of the decay behaviors of $N(1875)$ and $N(2080)$ is given.
Results obtained with typical cutoff values $\Lambda_0=1.0\ \mathrm{GeV}$ and $\Lambda_1=1.2\ \mathrm{GeV}$
are displayed in Table~\ref{table:total}.
\begin{table}[htpb]
\centering
\caption{\label{table:total}Partial decay widths of $N(1875)$ as $K \Sigma^*$ and $N(2080)$ as $K^*\Sigma$ molecule,
to different possible final states with $\Lambda_0=1.0\ \mathrm{GeV}$, $\Lambda_1=1.2\ \mathrm{GeV}$.
All of the decay widths are in the unit of $\mathrm{MeV}$, and the short bars denote that the corresponding
channel is closed or its contribution is negligible.}
\begin{tabular}{l*{3}{c}}
		\toprule\morecmidrules\toprule
		\multirow{3}*{Mode} & \multicolumn{3}{c}{Widths ($\mathrm{MeV}$)} \\
		\Xcline{2-4}{0.4pt}
		& \multicolumn{2}{c}{$J^P={3/2}^-$} & \multicolumn{1}{c}{$J^P={1/2}^-$} \\
		\Xcline{2-3}{0.4pt}\Xcline{4-4}{0.4pt}
		& $N(1875)$\ $K \Sigma^*$ & $N(2080)$\ $K^* \Sigma$ & $N(2080)$\ $K^* \Sigma$ \\
		\Xhline{0.8pt}
		$N\sigma(500)$ 	 		& 2.6  	 	 & 0.05		& 0.3   \\
		$\pi N$ 		 	    & 3.8    	 & 0.2  	& 22.7	\\
		$\rho N$  	     		& 2.3  	 	 & 3.8  	& 6.1	\\
		$\omega p$ 			 	& 6.6  	 	 & 11.3     & 18.2	\\
		$K\Sigma$ 		 	 	& 0.03   	 & 1.4  	& 9.1	\\
		$K\Lambda$ 		 	  	& 0.7    	 & 3.7   	& 19.3	\\
		$\eta p$ 			  	& 0.6   	 & 0.4    	& 1.8	\\
		$\pi\Delta$ 		 	& 201.4  	 & 82.6   	& 46.9	\\
		$K^*\Lambda$ 	  	 	& - 	 	 & 2.4    	& 7.9	\\
		$\phi p$ 	  	 		& -	     	 & 19.2    	& 27.0	\\
		$K\Sigma^*$ 	 		& -	   	 	 & 7.3    	& 1.3	\\
		$K\Lambda(1520)$ 	 	& -	   	   	 & 0.1    	& 1.3	\\
		$K\Lambda(1405)$ 	 	& -	   	 	 & 8.0    	& 8.8	\\
		$K\pi\Lambda$ 	 		& 10.1   	 & -    	& -		\\
		$K\pi\Sigma$ 	 		& -	   	 	 & 41.3    	& 46.1	\\
		\Xhline{0.8pt}
		Total 				 	& 228.2  	 & 181.7   	& 216.8	\\
\bottomrule\morecmidrules\bottomrule
\end{tabular}
\end{table}

\begin{figure}[htpb]
	\centering
        \includegraphics[width=0.49\textwidth]{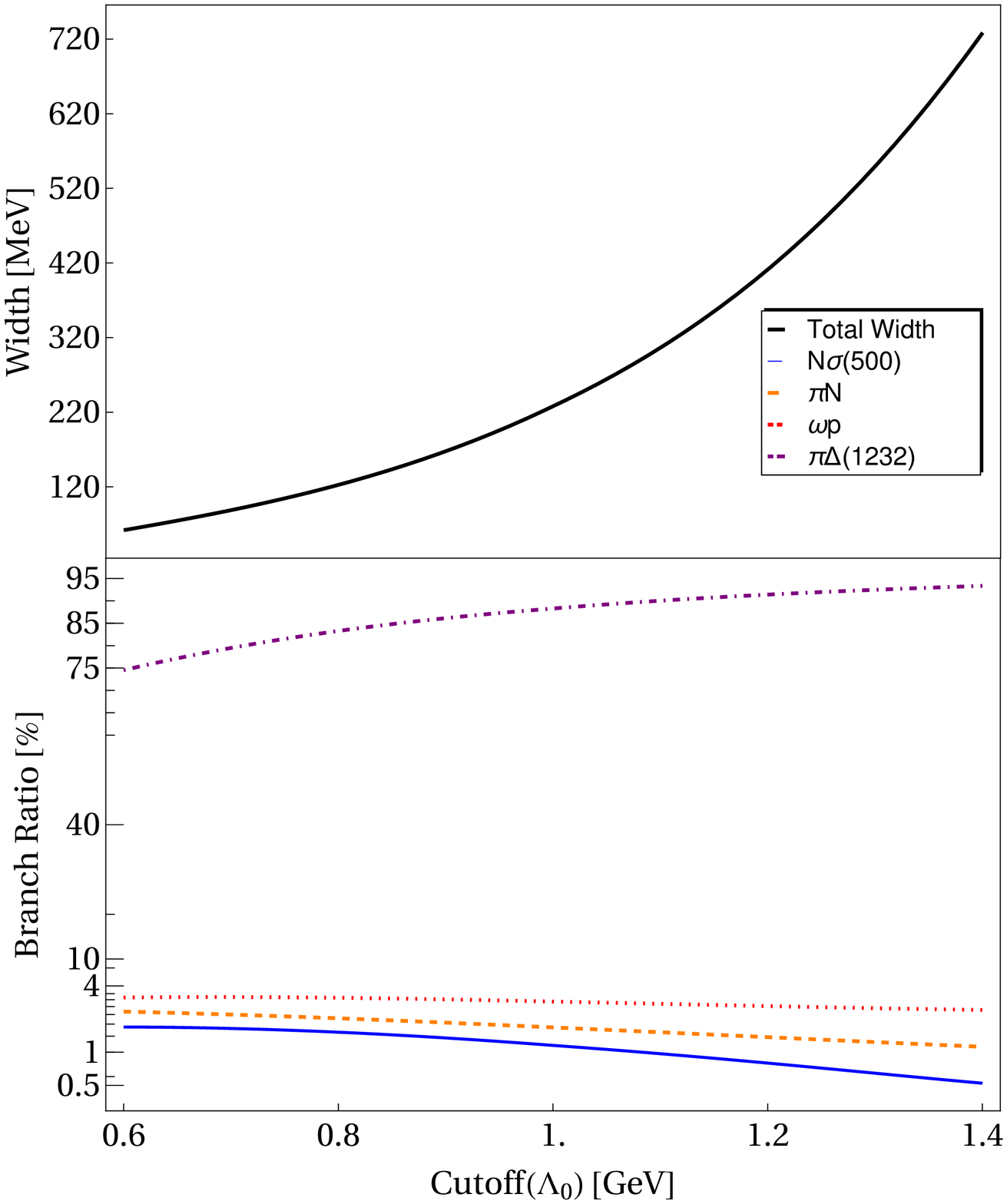}\hfill
        \includegraphics[width=0.49\textwidth]{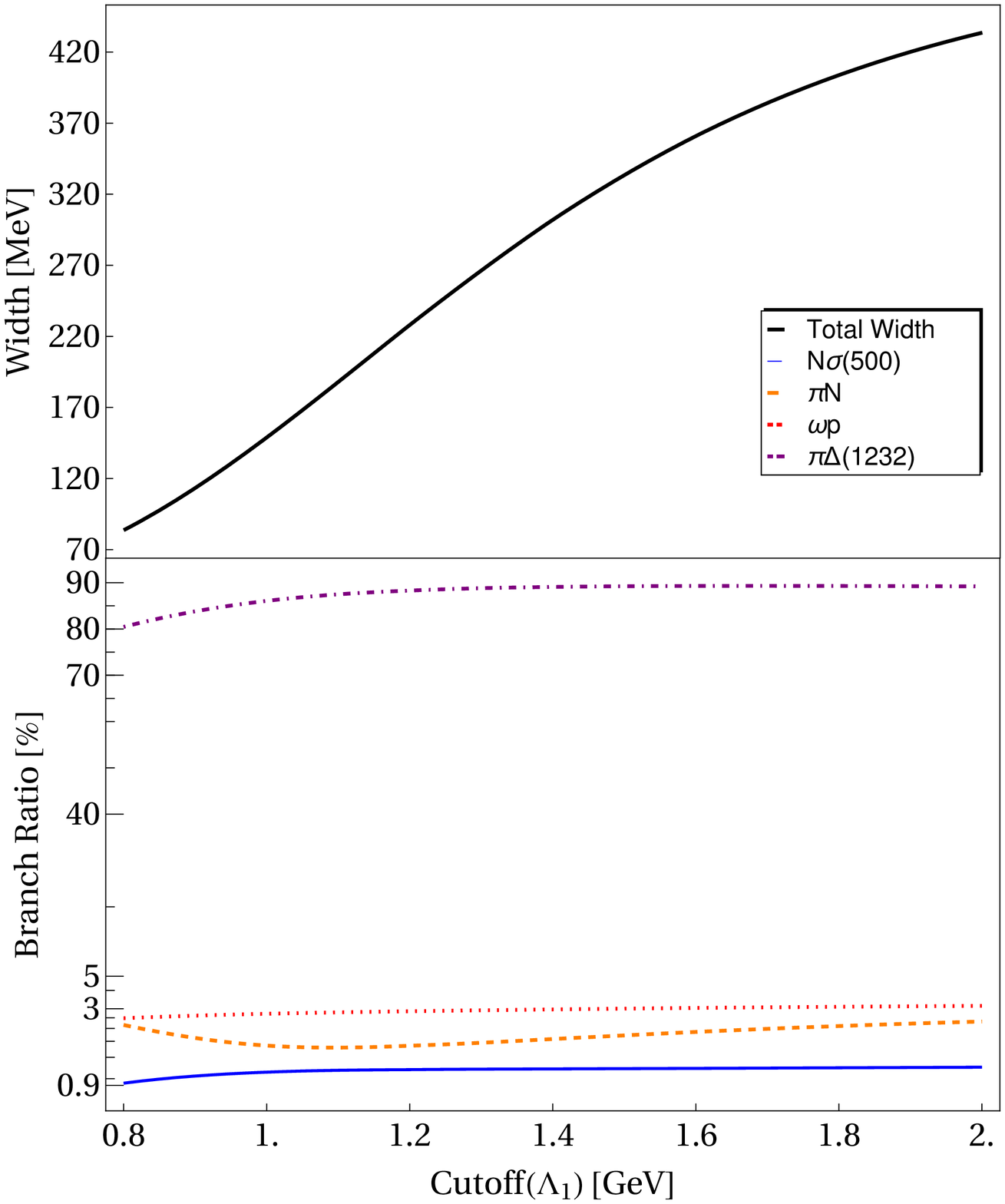}
\caption{\label{figure:dependence1}
Dependence of the total decay width and branching fractions of $N\sigma(500)$, $\pi N$, $\omega p$ and $\pi\Delta$ channels on the cutoffs in the $S$-wave $K\Sigma^*$ molecular scenario for $N(1875)$: (left) $\Lambda_0$ changes with $\Lambda_1$ fixed at $1.2\ \gev$; (right) $\Lambda_1$ changes with $\Lambda_0$ fixed at $1.0\ \gev$.}
\end{figure}

\begin{figure}[htpb]
	\centering
        \includegraphics[width=0.49\textwidth]{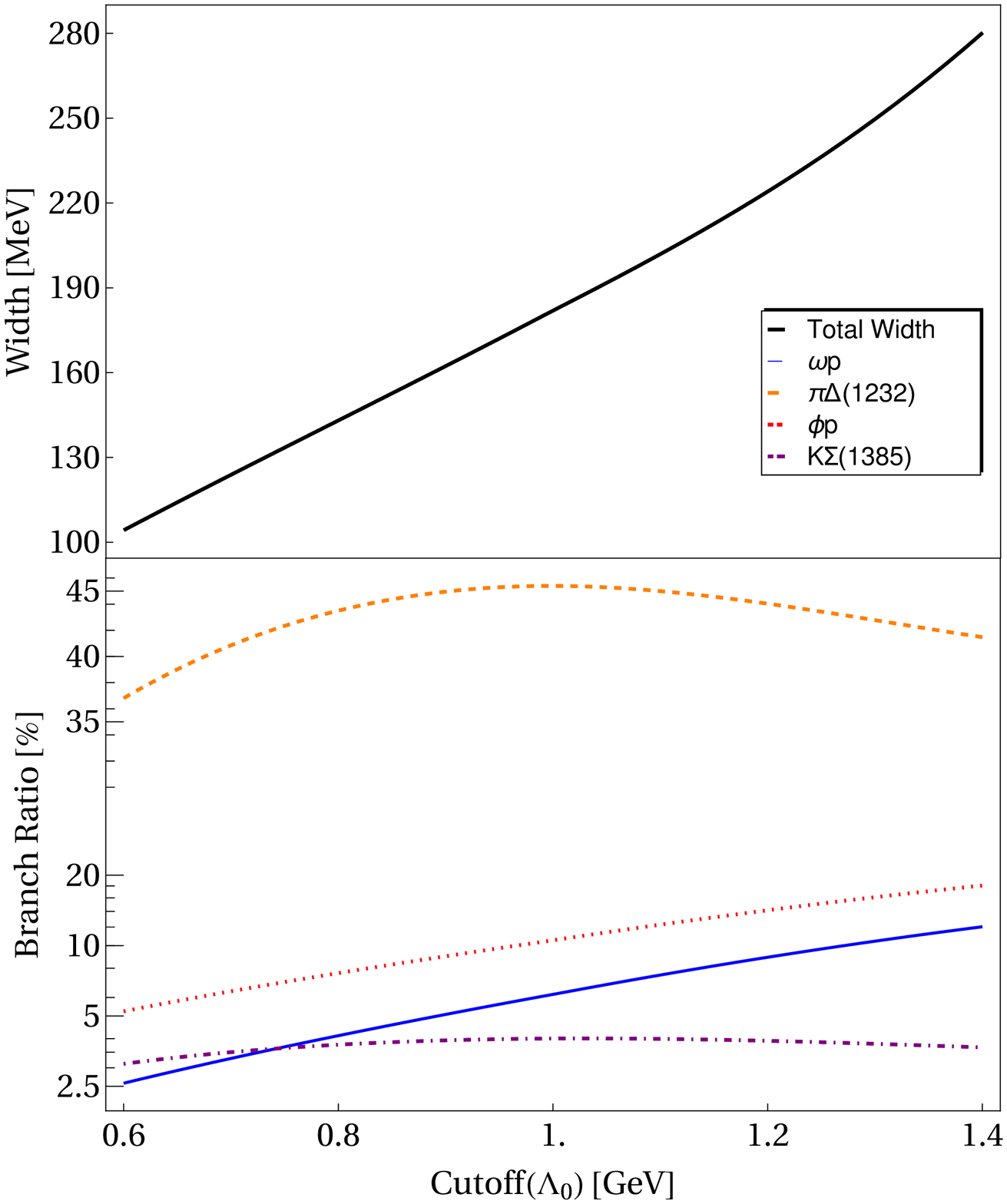}\hfill
        \includegraphics[width=0.49\textwidth]{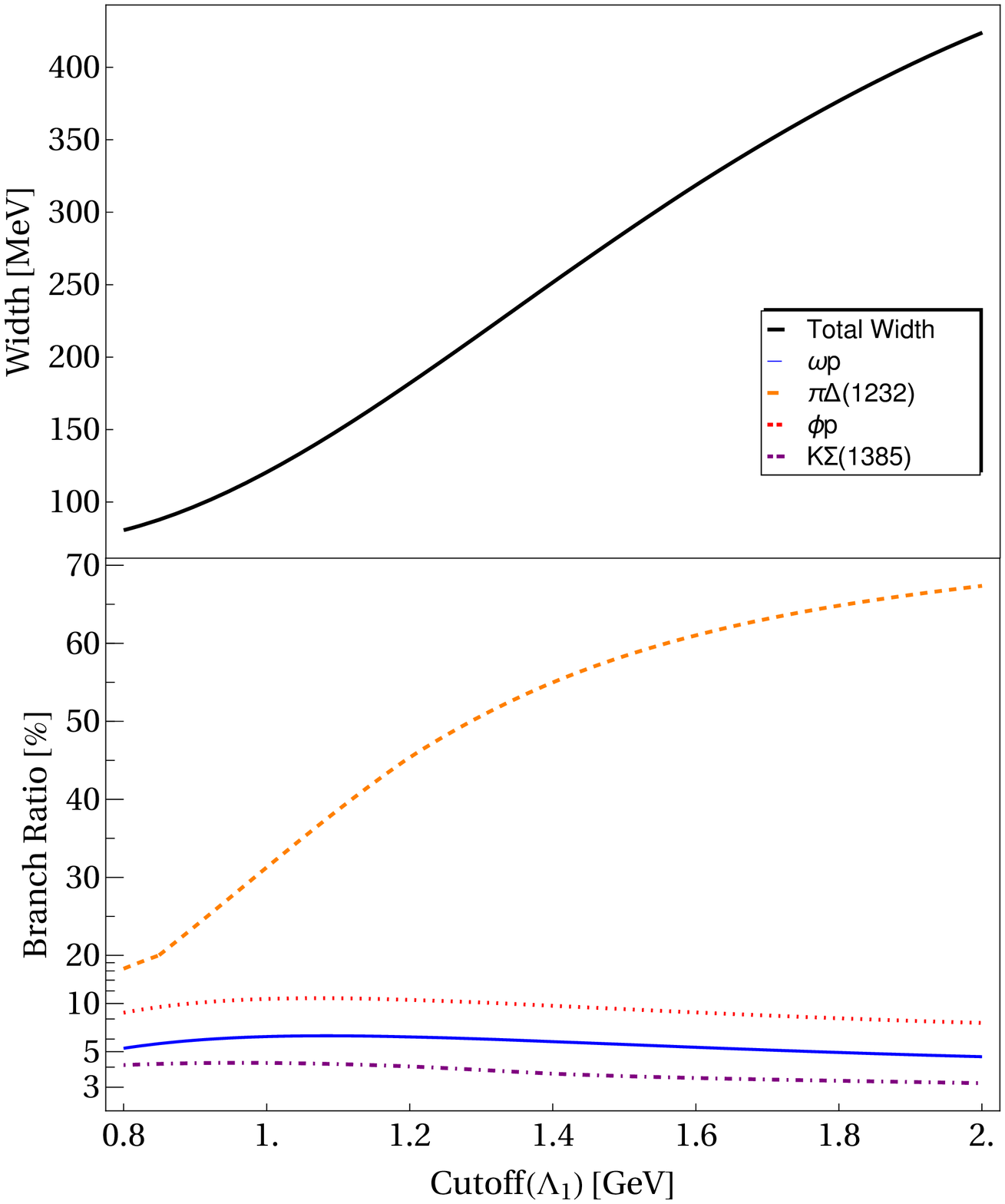}
	\caption{\label{figure:dependence2}
Dependence of the total decay width and branching fractions of $\omega p$, $\pi\Delta$, $\phi p$ and $K\Sigma^*$ channels on the cutoffs in the $S$-wave $K^*\Sigma$ molecular scenario for $N(2120)$ with spin-parity-${3/2}^-$: (left) $\Lambda_0$ changes with $\Lambda_1$ fixed at $1.2\ \gev$; (right) $\Lambda_1$ changes with $\Lambda_0$ fixed at $1.0\ \gev$.}
\end{figure}

The first nontrivial observation from Table~\ref{table:total} is that the
claimed total widths of $N(1875)$ and $N(2080)$ in $\mathrm{PDG}$ can be well reproduced in the $K \Sigma^*$ and $K^* \Sigma$ hadronic molecular pictures proposed in Refs.~\cite{He:2015yva,He:2017aps} where the $N(2080)$ was denoted as $N(2100)$.
The first four dominant two body decay channels are $\pi\Delta$, $\omega p$, $\pi N$ and $N\sigma(500)$ for $N(1875)$,
and $\pi\Delta$, $\phi p$, $\omega p$ and $K \Sigma^*$ for $N(2080)$ with spin-parity-${3/2}^-$.
The dependence of the total widths and the branching fractions of these dominant channels on the cutoffs are shown in Fig.~\ref{figure:dependence1} for $N(1875)$ and Fig.~\ref{figure:dependence2} for $N(2080)$. In $K \Sigma^*$ molecular scenario,  the branching fractions of the $\pi\Delta$, $\omega p$, $\pi N$ and $N\sigma(500)$ channels do not change much when the cutoffs increase and the $\pi\Delta$ channel contributes the dominant share of decay width to $N(1875)$ resonance in the whole ranges of both $\Lambda_0$ and $\Lambda_1$. It agrees with the results of multichannel partial-wave analysis in Ref.~\cite{Shrestha:2012ep}. However, the large contribution of the $N\sigma(500)$ channel for $N(1875)$ claimed in Refs.~\cite{He:2015yva,Sokhoyan:2015fra} do not show in our model.
Since both $\pi\Delta$ and $N\sigma$ lead to the same $N\pi\pi$ final state, we believe they have some cross-talk and suffer difficulty to be well separated experimentally. Note that the dependence of the total width on $\Lambda_0$ is stronger than its on $\Lambda_1$ in the $K \Sigma^*$ picture, while the case is inverse for the $K^* \Sigma$. It is because that the dominant contribution to the $\pi\Delta$ channel is the $K^*$ exchange for the $K \Sigma^*$ molecule and becomes $K$ exchange which is more dependent on the cutoff $\Lambda_1$ in the monopolar form factor Eq.~\eqref{eq:ff} for the $K^* \Sigma$ molecule. And it is also the reason why the branching fraction of the $\pi\Delta$ channel changed acutely when the cutoff $\Lambda_1$ increased in the right panel of Fig.~\ref{figure:dependence2}. As shown in Fig.~\ref{figure:dependence2}, $\pi\Delta$ is also the dominant decay channel of the $K^* \Sigma$ hadronic molecule. It coincides with the statements about $N(2080)$ in Ref.~\cite{Sokhoyan:2015fra}. Besides, the Refs.~\cite{Kiswandhi:2010ub,Ozaki:2009mj,Mibe:2005er} mentioned that the exotic $N^*$s with masses located around $2\ \gev$ have a significant contribution to the $\phi$-production. This property is well reflected in our numerical results that the partial width of $\phi p$ channel occupies a sizable part in the total decay width of $N(2080)$. Furthermore, our results also show a strong coupling of $N(2080)$ to $K\Sigma^*$ channel. It has been predicted in some very early works~\cite{Oh:2007jd,Capstick:1998uh}. In particular, the paper~\cite{Capstick:1998uh} claimed that for the spin-parity-${3/2}^-$ $N^*$ resonances the decay channels $K\Lambda(1405)$ and $K\Lambda(1520)$ are also opened and the width of $K\Lambda(1520)$ is smaller than the $K\Lambda(1405)$ channel. This is exactly verified by our results.

For the $S$-wave $K^* \Sigma$ molecular states, there is also a spin-parity-${1/2}^-$ one with its total width about 40 MeV larger than the spin-parity-${3/2}^-$ case. And the decay patterns of these two molecules are quite different. The most distinguishable property shown in the molecular pictures is that the spin-3/2 state has a very weak coupling to the $\pi N$ channel while the spin-1/2 state has a rather strong coupling to this channel. In fact, some evidence for such state from $\pi N$ experiment is listed under the two-star $N^*$ resonance $N(1895){1/2}^-$. And according to our calculation, it is possible to search for this spin-1/2 molecule in the $\pi N\to\phi N$ experiment.

In summary for the strange sector, the hadronic molecular scenario with $N(1875)$ and $N(2080)$ as $S$-wave $K\Sigma^*$ and $K^*\Sigma$ molecules, respectively, can provide a good explanation to their measured decay behaviors. This hadronic molecular scenario for the $N(1875)$ and $N(2080)$ is in fact also supported by the BESII data on $J\psi\to nK_S^0\bar\Lambda+c.c.$~\cite{Ablikim:2007ec} where three peaks are clearly visible in the $K_S^0\bar\Lambda$ invariant mass spectrum: one is near-threshold enhancement most probably due to $N(1535)$~\cite{Liu:2005pm}, other two peaks are around $N(1875)$ and $N(2080)$, respectively. With an order of magnitude more data on $J/\psi$ decays at BESIII now, a partial wave analysis of the $J\psi\to nK_S^0\bar\Lambda+c.c.$ data would be very valuable to pin down the properties of these three peaks. If the two higher peaks are indeed due to $N(1875)$ and $N(2080)$, one would also expect them appearing strongly in the $\pi\Delta$ invariant mass spectrum of $J/\psi\to \pi\Delta\bar p+c.c.$ data which should be checked by BESIII experiment. Another interesting channel to check is $J/\psi\to K\Sigma^*\bar p+c.c.$  to look for $N^*\to  K\Sigma^*$.  This scenario may also be checked by experiments at JLab and JPARC with $\gamma p$, $\pi p \to K\Lambda$, $K\Sigma^*$, $\phi p$, $\pi\Delta$, $\omega p$, etc.

\begin{figure}[htpb]
	\centering
        \includegraphics[width=0.49\textwidth]{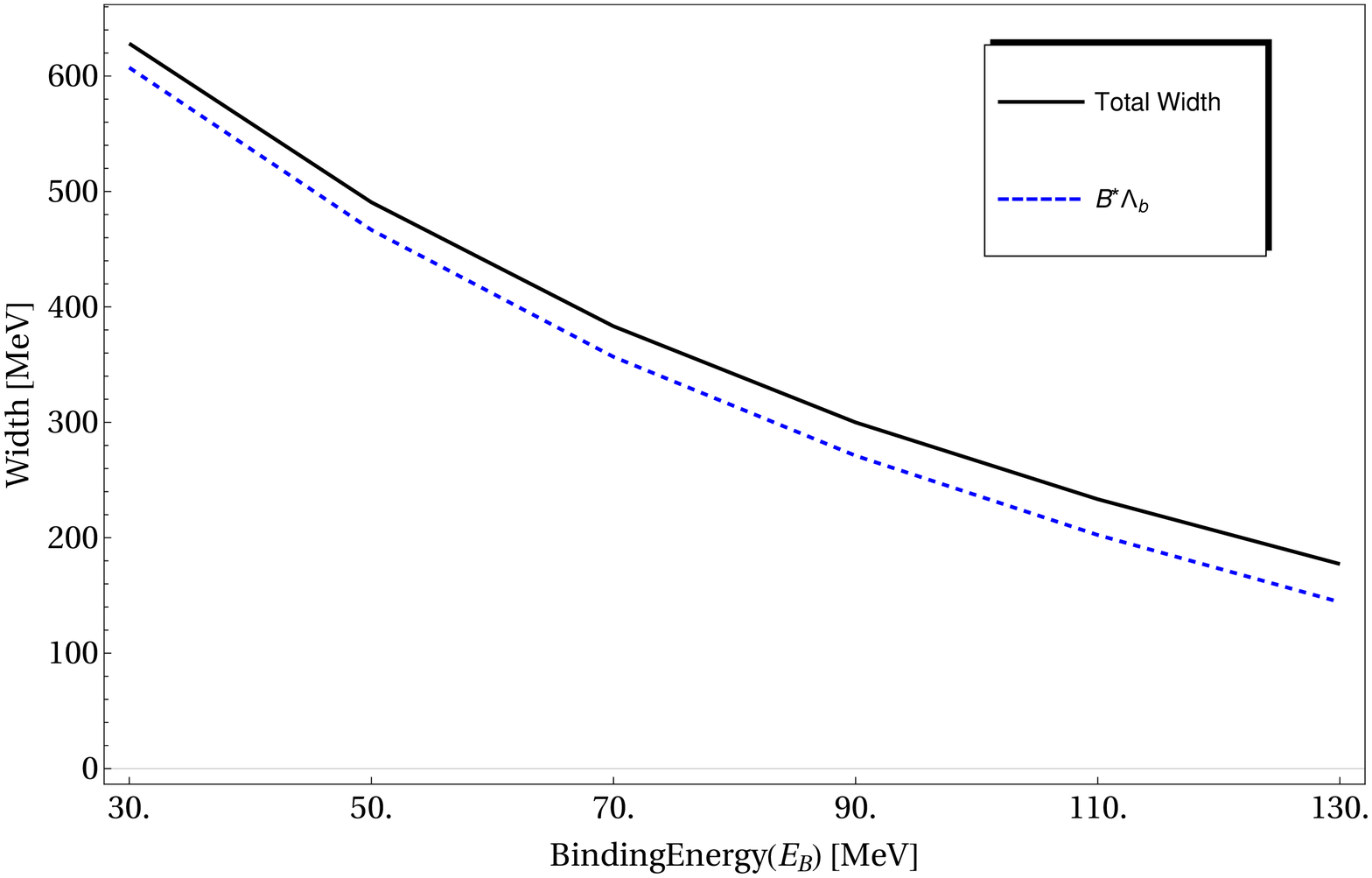}\hfill
        \includegraphics[width=0.49\textwidth]{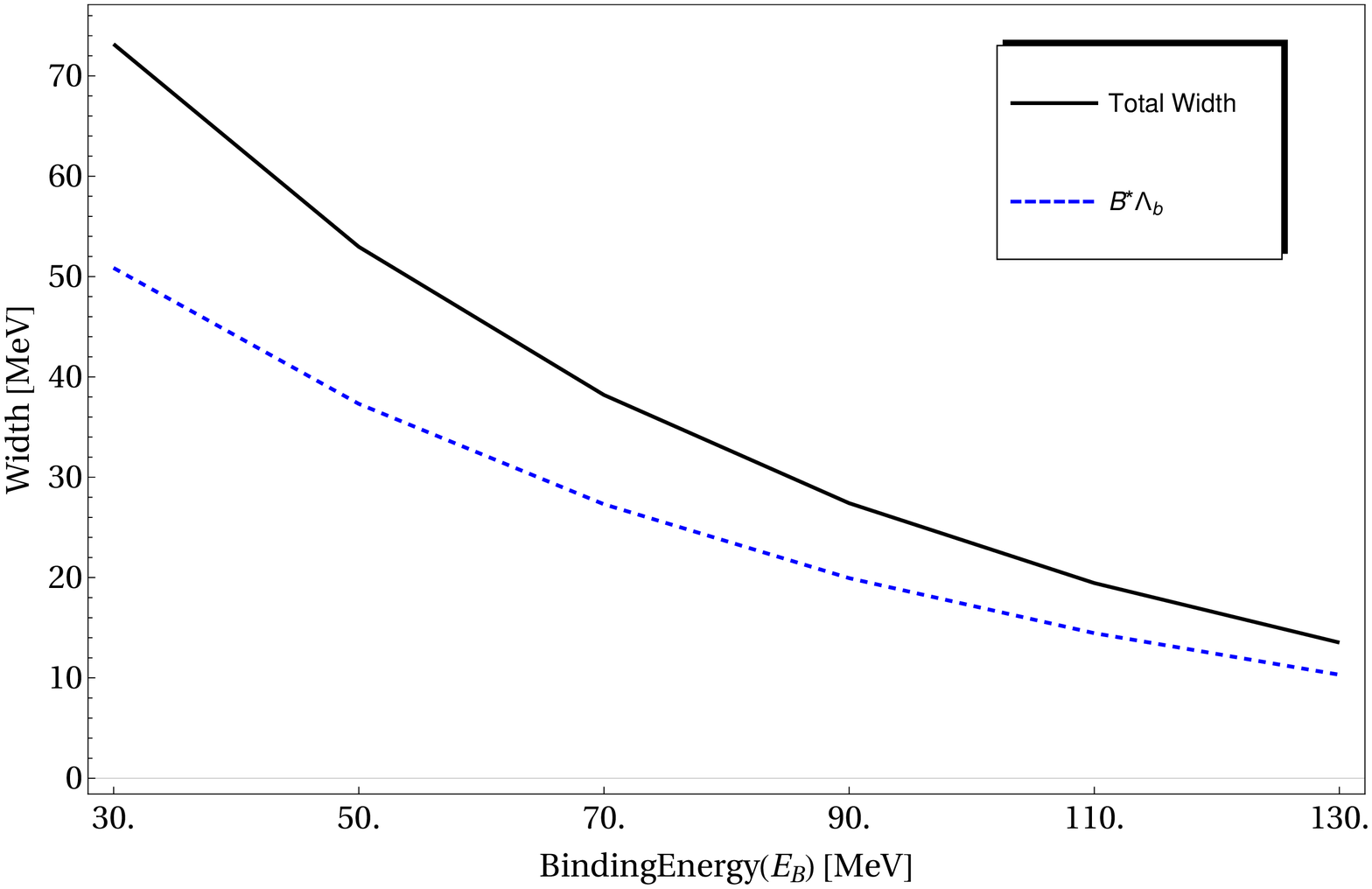}
	\caption{\label{figure:Eb}
Dependence of the total decay width (solid curve) and partial width (dashed curve) of $B^*\Lambda_b$ channel on the binding energy for the $B\Sigma_b^*$ (left) and $B^*\Sigma_b$ (right) molecules of spin-parity-${3/2}^-$.}
\end{figure}

With the exotic $N^*$s with hidden charm and hidden strangeness identified, more exotic $N^*$s with hidden beauty are expected. For the beauty sector, the binding energies for the lowest S-wave bound states range from 35 MeV~\cite{Wu:2010rv} to 130 MeV~\cite{Xiao:2013jla}. We take it as 90 MeV as obtained from a more dedicated recent coupled channel study for the $B\Sigma_b$ bound state~\cite{Shen:2017ayv}.  Then we obtain the partial decay widths for the $B \Sigma_b^*$ and $B^*\Sigma_b$ molecules as given in Table~\ref{table:total_B}. The largest partial decay width for both types of molecules is found to be $B^*\Lambda_b$ channel. The results are sensitive to the bind energy as shown in Fig.\ref{figure:Eb} for the two exotic $N^*$s with spin-parity-${3/2}^-$. 
Similar to the case for the $P_c$ states~\cite{Lin:2017mtz}, the pion exchange gives a large contribution to the decay width. The 
previous calculations~\cite{Wu:2010rv,Xiao:2013jla} have underestimated the decay width by only considering vector-meson exchanges.
For a thorough coupled channel study of these pentaquark states, the pion exchange is not negligible as also suggested by Refs.~\cite{Shimizu:2016rrd,Yamaguchi:2016ote,Shimizu:2018ran}. Nevertheless, all relevant studies support the existence of the pentaquark states with hidden beauty.
We expect future facilities, such as proposed electron-ion collider (EIC)~\cite{Accardi:2012qut} or EicC with center-of-mass energies of 12-30 GeV~\cite{EicC:2018}, to discover these very interesting exotic $N^*$s with hidden beauty.

\begin{table}[htpb]
	\centering
	\caption{\label{table:total_B}Partial widths of $B \Sigma_b^*$ and $B^*\Sigma_b$ molecule,
		to different possible final states with $\Lambda_0=1.0\ \mathrm{GeV}$, $\Lambda_1=2.0\ \mathrm{GeV}$.
		All of the decay widths are in the unit of $\mathrm{MeV}$, and the short bars denote that the corresponding
		channel is closed or its contribution is negligible. The binding energy of these two molecules are set as 90.0 $\ \mathrm{MeV}$.}
	\begin{tabular}{l*{3}{c}}
		\toprule\morecmidrules\toprule
		\multirow{3}*{Mode} & \multicolumn{3}{c}{Widths ($\mathrm{MeV}$)} \\
		\Xcline{2-4}{0.4pt}
		& \multicolumn{2}{c}{$J^P={3/2}^-$} & \multicolumn{1}{c}{$J^P={1/2}^-$} \\
		\Xcline{2-3}{0.4pt}\Xcline{4-4}{0.4pt}
		& $B \Sigma_b^*$ & $B^* \Sigma_b$ & $B^* \Sigma_b$ \\
		\Xhline{0.8pt}
		$B^*\Lambda_b$ 	 		& 271.1  	 & 19.9		& 167.0 \\
		$\Upsilon p$ 		 	& 0.3    	 & 0.04  	& 0.1	\\
		$\rho N$  	     		& 5.5  	 	 & 0.02  	& 0.1	\\
		$\omega p$ 			 	& 20.9  	 & 0.07     & 0.4	\\
		$B\Lambda_b$ 		 	& -   	 	 & 7.3  	& 135.9	\\
		$B\Sigma_b$ 		 	& -    	 	 & -   		& -		\\
		$\eta_b p$ 			  	& 0.02   	 & 0.0001   & 0.0009	\\
		$\chi_{b0} p$ 		 	& 1.4  	 	 & 0.0008   & 0.2	\\
		$\pi N$ 	  	 		& 0.7 	 	 & 0.005    & 0.003	\\
		$B\Sigma_b^*$ 	  	 	& -	     	 & -    	& -		\\
		\Xhline{0.8pt}
		Total 				 	& 299.9  	 & 27.4   	& 303.8	\\
		\bottomrule\morecmidrules\bottomrule
	\end{tabular}
\end{table}

\bigskip

\section*{Acknowledgments}

We thank Feng-Kun Guo and Jun He for helpful discussions. This project is supported by NSFC under Grant No.~11621131001 (CRC110 cofunded by DFG and NSFC) and Grant No.~11747601.



\end{document}